\documentclass{article}

\usepackage{arxiv}

\usepackage[utf8]{inputenc} 
\usepackage[T1]{fontenc}    
\usepackage{hyperref}       
\usepackage{url}            
\usepackage{booktabs}       
\usepackage{amsfonts}       
\usepackage{nicefrac}       
\usepackage{microtype}      
\usepackage{lipsum}		
\usepackage{graphicx}
\usepackage[square,sort,comma,numbers]{natbib}
\usepackage{doi}
\usepackage{titlesec}
\usepackage{enumitem}
\usepackage{hyperref}

\usepackage{todonotes}

\title{Towards Civic Digital Twins: Co-Design the Citizen-Centric Future of Bologna}

\author{Massimiliano Luca\\
	Mobile and Social Computing Lab\\
	Fondazione Bruno Kessler\\
	Trento, 38100, Italy\\
	\texttt{mluca@fbk.eu} \\
	\And
    Bruno Lepri\\
	Mobile and Social Computing Lab\\
	Fondazione Bruno Kessler\\
	Trento, 38100, Italy\\
	\texttt{lepri@fbk.eu} \\
	\And
    Riccardo Gallotti\\
	Complex Human Behaviour Lab\\
	Fondazione Bruno Kessler\\
	Trento, 38100, Italy\\
	\texttt{rgallotti@fbk.eu} \\
	\And
    Stefania Paolazzi\\
        Strategic projects City of Knowledge and Digital Twin\\
        Comune di Bologna\\
        Bologna, 40100, taly\\
	\texttt{stefania.paolazzi@comune.bologna.it} \\
	\And
    Mauro Bigi\\
	Research-Action and Development Unit\\
        Fondazione IU Rusconi Ghigi\\
	Bologna, 40100, taly\\
	\texttt{mauro.bigi@fondazioneinnovazioneurbana.it} \\
	\And
    Marco Pistore\\
        Modeling and Simulation of Socio-Technical Systems\\
	Fondazione Bruno Kessler\\
	Trento, 38100, Italy \\
	\texttt{pistore@fbk.eu} 
 }

\renewcommand{\headeright}{Preprint}
\renewcommand{\undertitle}{Preprint}
\renewcommand{\shorttitle}{Towards Civic Digital Twin}

\begin{document}
\maketitle

\begin{abstract}
We introduce \emph{Civic Digital Twin (CDT)}, an evolution of Urban Digital Twins designed to support a citizen-centric transformative approach to urban planning and governance. CDT is being developed in the scope of the Bologna Digital Twin initiative, launched one year ago by the city of Bologna, to fulfill the city’s political and strategic goal of adopting innovative digital tools to support decision-making and civic engagement.
The CDT, in addition to its capability of sensing the city through spatial, temporal, and social data, must be able to model and simulate social dynamics in a city: the behavior, attitude, and preference of citizens and collectives and how they impact city life and transform transformation processes. Another distinctive feature of CDT is that it must be able to engage citizens (individuals, collectives, and organized civil society) and other civic stakeholders (utilities, economic actors, third sector) interested in co-designing the future of the city.
In this paper, we discuss the motivations that led to the definition of the CDT, define its modeling aspects and key research challenges, and illustrate its intended use with two use cases in urban mobility and urban development.
\end{abstract}


\section{Motivation}

The amount of data available to sense cities is growing rapidly, and their spatial, social, and temporal resolution has become more fine-grained than ever before \cite{zheng2019urban}. Computational capabilities also increased quickly, allowing cities to consider new paradigms to enable data-driven decision-making. 
One of the emerging paradigms in this sense is Urban Digital Twins (UDTs). UDTs are 
the virtual representation of a city's physical assets, processes, and systems, exploiting data and data analytics to help build simulation models and support decision processes \cite{bettencourt2024recent,caldarelli2023role}. Cities worldwide aim to build their UDTs to monitor different aspects of cities. Examples are UDTs for traffic management in Zurich \cite{schrotter2020digital}, for road infrastructures in Tokyo \cite{el2020roads}, for water management in Odense \cite{pedersen2021living}, for geographical modeling in Vienna \cite{lehner2020digital}, for 3D urban modeling in Singapore and Helsinki \cite{helsinki2021}, for service accessibility in Barcelona \cite{barcelona2022}, and for many other aspects \cite{faliagka2024trends, jafari2023review}. 
Our commentary focuses on the city of Bologna's UDT initiative.

Bologna is Italy's seventh-largest metropolitan area, with almost one million people living there. Inevitably, Bologna faces the same societal challenges as many other urban agglomerations, such as traffic and pollution management, energy management, and reducing inequalities. At the same time, Bologna is engaged in a very ambitious strategy for the social, environmental, and technological transition of the city: besides addressing the climate challenge -- Bologna is one of the over 100 cities committed to achieving climate neutrality by 2030 -- the strategy aims to shift the city’s economic and social development towards a knowledge-based dimension, involving the regeneration of a substantial urban area and the implementation of active policies.

In Bologna, a UDT initiative was launched in 2023 to fulfill the city's political and strategic goal, namely as a key instrument to support decision-making in tackling urban challenges and developing city transformation strategies. The goal is to leverage the methodologies and results of other cities that have already developed UDT while embracing the ambitious challenge of providing the city of Bologna with a UDT focused on the civic dimension of the city: a Civic Digital Twin (CDT). 
\emph{The biggest ambition of our project and what differentiates a CDT from other UDTs is that, in addition to modeling and simulating the physical city in a virtual environment, we want our digital twin to model and simulate people's behaviors, attitudes, and choices.} 
This aspect is of crucial importance when analyzing urban transformations: the impact and outcomes of these transformations highly depend on citizens’ individual and collective attitudes towards change, their engagement and collaboration in -- or their opposition to -- the transformation processes, the conflictual dynamics among citizens and collectives, as well as their intentional or accidental misuse of the instruments and incentives implemented by the city administration.
A must-be-solved related challenge that we are aiming to tackle with our CDT is to model \emph{cities as systems-of-systems} by leveraging the advantages that complex systems can bring to DTs \cite{bettencourt2024recent,caldarelli2023role} and by exploiting these approaches to model complex civic behaviors. Indeed, as most of the existing UDTs aim to simulate specific problems individually, cities are complex ecosystems, and single decisions on specific tasks, like road closures to reduce pollution, may profoundly alter other dynamics, such as economic growth or segregation of a city district \cite{mohl1993race}. Modeling such aspects and interplays is fundamental in CDT to make informed decisions by being aware of the potential side-effects of decisions and the level of uncertainty that complex dynamics may impose on the outcomes.

Bologna CDT also has another ambition, which we think is strongly intertwined with the previous one: \emph{ensuring that the CDT is not a technical tool in the hands of public offices but a civic tool available to all citizens}, supporting as a citizen-in-the-loop approach that spaces, from co-designing of the CDT solution to engagement in the definition of the urban transformation goals and strategies to active participation in the decision making and then in the actuation and monitoring of the transformation process. 
The interesting aspect of a CDT is that citizens play a crucial role both within the model and as active participants during the co-design phase and the progress monitoring of the urban transformations. 
While there are already some conceptual models to build digital twins that enable community feedback \cite{white2021digital} or that involve citizens in the problem-definition phase \cite{dembski2020urban} and, in general, in city planning \cite{batty2024digital}, Bologna's CDT will be the first twin in which citizens will be involved with such a prime actor role.

Designing and implementing CDTs is a long and complicated journey that presents various challenges. 
The commitment of the city of Bologna, as well as the recent progress in data availability, computational power, and urban modeling and simulation approaches, make us confident that the Bologna CDT has an opportunity to become a successful innovative tool for city decision-making and that the path to this result will produce important scientific results.
This paper aims to share the goals and requirements of a CDT, as we understand them after one year of the project. 

\section{Modeling the Civic Digital Twin}

\begin{figure}
    \centering
    \includegraphics[width=0.9\linewidth]{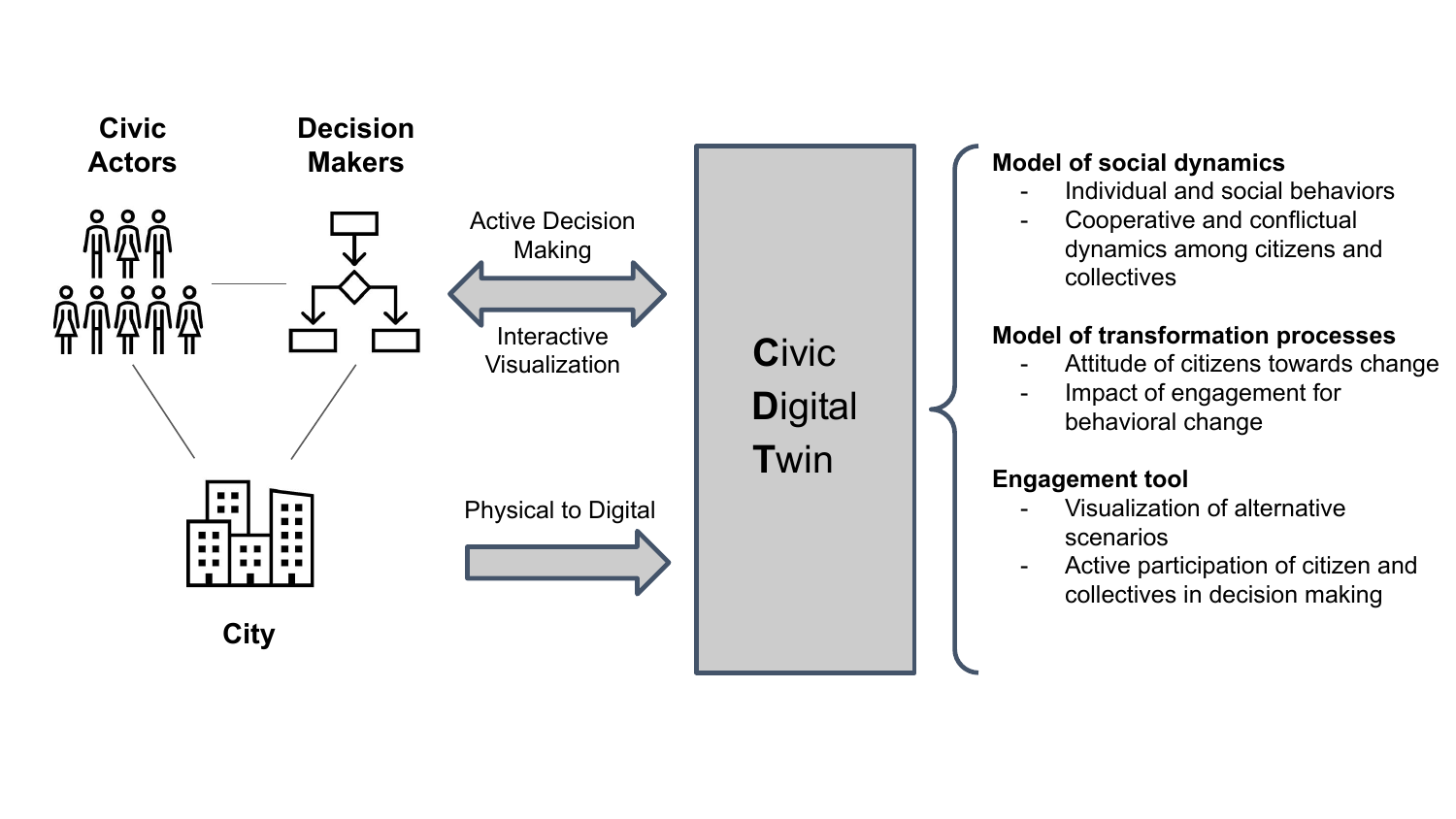}
    \caption{Civic Digital Twin: Conceptual representation of the model.}
    \label{fig:cdt}
\end{figure}

To fulfill the ambitious objectives just described, we imagined a CDT as a framework able to model multi-layered urban processes integrating human behaviors, societal interactions, and environmental dynamics. 
In particular, a CDT operates across three interconnected domains: \emph{i)} accurate modeling of social dynamics; \emph{ii)} analyze, simulate, and predict urban transformation processes; and \emph{iii)} effective citizen engagement: see Figure~\ref{fig:cdt}.

Cities are characterized by a heavily built environment and high population density. The physical layer -- i.e., land morphology and the built environment -- and the social layers -- i.e., citizens, their behaviors and processes, their organizations and norms -- influence, limit, and transform each other.   
The CDT is a \emph{model of social dynamics}, that is, it focuses on the social layer: it models the behaviors and interactions of citizens at both individual and collective levels, capturing the nuances of cooperative and conflictual interactions among various social groups. By understanding how citizens interact with the physical layer, organize their behaviors into more or less structured processes, respond to urban challenges, and interact within communities, the CDT can analyze the city's current functioning and simulate the effects of policies or interventions. 
For example, it can predict the social implications of introducing a new public transport system or regenerating a city district, revealing opportunities for optimization and unintended consequences, including their impact on other city systems, and explaining the influence of a cooperative or conflictual attitude of individuals and collectives towards the new policy. 
The CDT is hence also a \emph{model of transformation processes}: the need for sustainable development, technological innovation, and societal well-being drives the evolution of cities; the CDT delves into transformation processes by analyzing their adoption and long-term effects, focusing on the changes on social dynamics that are enacted by these processes, on citizen attitudes toward change, and on their behavioral shifts. 

We have already remarked that citizens' attitudes are of fundamental importance in urban processes, particularly in urban transformation processes, hence the importance of citizen engagement that, facilitated by digital technologies, can help shape these transformation processes \cite{helbing2024co}.

Thus, in addition to a modeling framework, the CDT wants to serve as \emph{tool for citizen engagement}. 
The CDT is devoted to creating a bridge between abstract urban models and everyday citizen experiences. Providing intuitive, interactive interfaces empowers individuals and collectives to visualize how proposed changes affect their neighborhoods, commute times, or access to resources.
By allowing citizens to understand the expected outcomes of policies and transformation processes, as well as to compare them with different scenarios, the CDT reinforces and empowers the active role of citizens and stakeholders in city governance, both when policies have to be deliberated and when their progresses, effects and impacts have to be monitored and assessed.  
This may lead to deeper trust and collaboration between citizens and decision-makers, help align diverse perspectives into a shared vision for the city, and democratize the decision-making process, with important implications for the speed and outcome of transformation processes.

We designed the CDT as a dynamic, evolving framework that thrives on iterative feedback. Outputs generated by the CDT -- such as policy simulations, urban development scenarios, or predicted behavioral outcomes -- are shared with citizens, policymakers, and urban planners in a continuous loop. Citizens can use such a feedback loop: outputs allow individuals to understand the broader implications of their preferences and actions, inspiring informed participation in shaping their urban environment. Similarly, decision-makers receive precise, scenario-based analyses highlighting potential trade-offs, conflicts, or synergies in urban development strategies.

\section{Adopting the CDT: Two Use Cases}
Here, we describe two use cases related to mobility and urban development to show how a CDT may act.
In both use cases, the tackled problems are relevant for the future development of the city of Bologna, and for cities in general, decision-making is difficult due to both the complexity of the phenomena to be analyzed and the role of citizens and their attitude, and -- we argue -- the CDT can offer invaluable support.

\subsection{Mobility: Understanding the impact of traffic regulations}

Regulating and limiting the access of private vehicles into growing city areas is a policy adopted in Italy (e.g., Milan), Europe (e.g., London and Stockholm), and worldwide (e.g., Luoyang and Shenzhen in China) to reduce traffic and hence lower the emission of pollutants and climate-altering gasses. This policy can also have other side effects, such as reallocating urban space from roads and parking areas to other usages, incentivizing soft mobility and public transportation, and so on. It also has potential side effects, both in the social domain (e.g., limiting the mobility possibilities of vulnerable groups) and in the economic domain (e.g., affecting access to city centers, tourism, and commerce).
Behavioral change in response to access regulations plays a central role in this scenario: the impact on traffic and pollutants depends on the rigidity of people's behaviors, and also, if behaviors change, shifts in the time cars enter the city have a different impact than modal shifts.
Also, traffic limitations are a controversial topic, and the attitude of people toward this change can play an important role in its effects, hence the opportunity for information and engagement campaigns. 

In this context, the CDT can support decision-making on this complex topic, helping to understand and explain the impacts of different regulations (e.g., their territorial extension, time extension, affected vehicles, access costs, etc.) in terms of citizen behaviors (rigid habits, time shifts, modal shifts, reduced number of accesses, etc.), traffic and emissions, as well as to link all of this to indirect social and economic impacts. This requires integrating different models: of city traffic, of human behavior in reaction to tariffs, of civic acceptance, and of social and economic implications. Also, the CDT can help illustrate and explain the different scenarios to decision-makers and citizens, drive the decision-making process, explain different behavioral change patterns that can lead to different outcomes, and increase the level of awareness and participation.

\subsection{Urban planning: Active engagement of city neighborhoods}

Districts can play an important role in urban planning: city districts are indeed very heterogeneous, differing for land morphology, centrality, available infrastructures and services, demographic, social, and economic aspects, and so on. This heterogeneity may be a competitive advantage for cities but may also be the source of underdevelopment and social inequalities. The active engagement of neighborhoods in urban planning has the goal of helping the city achieve harmonic development, ensuring that each neighborhood can evolve according to its strengths and vocations, reducing at the same time inequalities and conflicts.

Achieving this goal requires analyzing the current situation of city districts by integrating city statistical data, city sensing (e.g., sensing of daily routines of citizens and neighborhoods), and sentiment analysis; the goal is to understand the strengths and weaknesses of the districts, both in absolute and comparative terms. This "as-is" situation of the city districts needs to be complemented with an analysis of the districts "to-be": this requires understanding the expectations and the vocation of the citizens in the different neighborhoods through processes of listening, dialogue, and collaboration in each neighborhood to reveal priorities, needs, indications, and solution proposals. 
Finally, the actions of urban plans need to be directed to the priorities that emerged from this analysis and collaboration process.

The CDT can help this process, both in the "as-is" and "to-be" analysis, by helping to understand the complex interactions among all the different dimensions that characterize a neighborhood, including the expectations of its citizens.
It can also help as a communication and engagement tool in the "on-the-field" activities with the neighborhoods by visualizing the districts' status and the different possible evolution trajectories, aligning diverse perspectives, and supporting the emergence of a shared vision for the neighborhood.

\section{Conclusions}

In this paper, we have introduced Civic Digital Twins, an innovative development of Urban Digital Twins aiming at better capturing a key aspect of cities -- the role of citizens in urban dynamics and urban transformation processes -- and at reinforcing the active role of citizens and stakeholders in city governance.

We are currently engaged in addressing the research challenges associated with CDT. They concern the modeling, analysis, and prediction capabilities, where the key aspects are the \emph{complexity} of the system under analysis and the \emph{uncertainty} in the outcomes of analysis and prediction tasks. They also concern the use of CDT for civic engagement, where the key aspects are \emph{explainability}, namely the ability to present the results of CDT analysis and prediction tasks in a way that "makes sense" to citizens and decision-makers, and \emph{accountability}, namely the ability to connect decisions to expected objectives, to monitor the progress towards these expected outcomes, and to support audit and revision processes. 

In the context of the city of Bologna, the development of CDT will progress in parallel to its adoption on problems of increasing complexity and impact, thus permitting an incremental research-development-validation loop and progressive adoption of CDT as a decision-making tool for tackling urban challenges and developing city transformation strategies.

\bibliographystyle{plain}
\bibliography{references} 

\end{document}